\documentclass[showpacs,twocolumn,pre,floatfix]{revtex4}
\usepackage{psfrag,epsfig,amsfonts,amssymb,amsmath,wasysym}
\usepackage{dcolumn}

\usepackage[normalem]{ulem}
\usepackage{color}

\begin{document} 
\newcommand{\RR}{{\mathbb R}}
\newcommand{\pos}{x}
\newcommand{\cm}{X}
\newcommand{\ori}{\phi}
\newcommand{\poscm}{y}
\newcommand{\per}{L}
\newcommand{\kB}{k}
\newcommand{\fluc}{\zeta}
\newcommand{\uint}{W}

\newcommand{\abs}[1]{\vert #1 \vert}
\newcommand{\cyan}{cyan}
\newcommand{\Cyan}{Cyan}

\title{Dimer motion on a periodic substrate:
Spontaneous symmetry breaking and absolute negative mobility}

\author{David Speer$^{1}$}
\author{Ralf Eichhorn$^{2}$}
\author{Mykhaylo Evstigneev$^{1}$}
\author{Peter Reimann$^{1}$}
\affiliation{$^{1}$Universit\"at Bielefeld, Fakult\"at f\"ur Physik, 33615 Bielefeld, Germany
\\
$^{2}$NORDITA, Roslagstullsbacken 23, 10691 Stockholm, Sweden}

\begin{abstract}
We consider two coupled particles moving 
along a periodic substrate potential 
with negligible inertia effects (overdamped limit).
Even when the particles are identical and the 
substrate spatially symmetric, a sinusoidal external 
driving of appropriate amplitude and frequency
may lead to spontaneous symmetry breaking in the
form of a permanent directed motion of the dimer.
Thermal noise restores ergodicity and thus
zero net velocity, but entails arbitrarily fast 
diffusion of the dimer for sufficiently weak noise.
Moreover, upon application of a static bias force, 
the dimer exhibits a motion opposite to 
that force (absolute negative mobility).
The key requirement for all these effects 
is a non-convex interaction potential 
of the two particles.
\end{abstract}


\maketitle

\section{Introduction}
The dynamics of a dimer in the presence of
a periodic potential is of interest in a 
variety of different contexts.
A first example is a diatomic 
molecule adsorbed and moving on a crystal surface 
\cite{boisvert97,qin00,pijper07},
e.g. during crystal growth or thin film formation
\cite{braun03,fusco03,romero04,heinsalu08}.
In many cases, the dimer is able to explore the
entire two-dimensional surface, but in some cases 
also an effectively one-dimensional motion
arises, e.g. due to a strong crystaline anisotropy,
a pronounced step, or a channeled (110) metal surface
\cite{fusco03,romero04}.
As an example, Fig. 1 illustrates a Si$_2$ or Ge$_2$ 
dimer confined between the rows 
of the reconstructed Si(001) or Ge(001) surface, 
see \cite{pijper05} and references therein.
Motion of two coupled atoms (or defects, nanoparticles, etc.)
in a three-dimensional (atomic, optical etc.) 
lattice is yet another option along these lines.
Further important examples are simple 
models of processive molecular motors
consisting of two elastically coupled ``heads'' (motor domains) 
and walking along a polymer filament 
\cite{ajd94,pes95,derenyi96,julicher97,mog98,ast99,str99,li00,els00,klumpp01}.
Artificially manufactured set-ups include 
colloidal doublets \cite{tie08} in
periodic arrays of magnetic \cite{magnetic}
or optical \cite{optic} traps.
Last but not least, dimer models are 
well established to describe the sliding friction 
of two coupled asperities at the nanoscale
\cite{goncalves04,maier05,goncalves05,tiwari08},
but also arise in less ``obvious'' systems
like superionic conductors \cite{vollmer79},
dissociated dislocations \cite{patriarca05},
and SQUIDs \cite{squids,jan11}.

\begin{figure}[h] 
\epsfxsize=0.75\columnwidth
\epsfbox{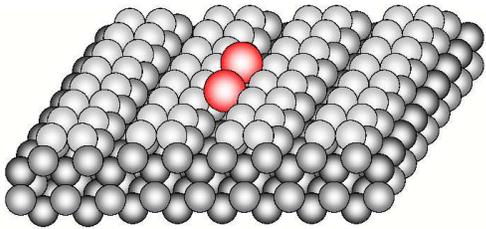} 
\caption{(Color online):
Schematic illustration of the confined motion
of a Si$_2$ or Ge$_2$ dimer (big red spheres) within the rows 
of a 2$\times$1 reconstructed
Si(001) or Ge(001) surface (small grey spheres).
}
\label{fig1}
\end{figure}

In most of the previous theoretical studies, the dimer was 
driven out of equilibrium by some external 
forcing and the spatial inversion symmetry of the 
system was broken a priori in one or another way.
Most prominently, in the context of so called
ratchet effects, a spatial anisotropy of
either the substrate potential
or of the dimer itself was assumed 
\cite{ajd94,derenyi96,dia96,dia97,cil98,cil01,mog98,str99,els00,li00,klumpp01,dan03,%
mateos04,craig06,retkute06,menche06,kuma08,mateos08,gehlen08,pototsky09,%
makhnovskii09,gehlen09,pototsky10,fornes10,vincent10}.
Another important direction concerns the behavior of 
drift and diffusion when the spatial symmetry is 
broken by an externally applied, constant ``bias''-force
\cite{goncalves05,fusco05,patriarca05,heinsalu08,evstigneev09}.
Studies of fully symmetric dimer models have focused on
the diffusive properties, since a systematic directed
dimer motion seems -- at first glance ``obviously'' -- 
not possible in such systems
\cite{boisvert97,braun03,goncalves04,romero04,pijper05,pijper07,bammert08}.
Exceptions are some idealized conservative (Hamiltonian) 
models \cite{fusco03,hennig08} 
and the recent paper \cite{mulhern11}, 
to which we come back in Sec.~X.

In our present study, the main focus is on 
dimers consisting of identical particles 
and on spatially periodic systems without
any intrinsically broken symmetry.
To avoid unnecessary complications, we furthermore
assume that inertia effects are negligible (overdamped limit)
\cite{ajd94,derenyi96,dia96,dia97,cil98,cil01,mog98,str99,els00,li00,klumpp01,dan03,%
mateos04,retkute06,menche06,mateos08,gehlen08,pototsky09,%
gehlen09,pototsky10,fornes10,evstigneev09},
and that the dimer is driven out of equilibrium by a 
sinusoidally oscillating force 
\cite{dia97,cil98,cil01,denisov02,mateos04,menche06,fornes10,mulhern11,vincent10}.
Our first main result will be that -- in 
spite of the spatial symmetry of the system -- 
the deterministic dynamics still admits a
systematic net motion of the dimer via
spontaneous symmetry breaking.
In particular, ergodicity is broken,
and thus the direction of the spontaneous 
transport depends on the initial conditions.
Upon including thermal noise, ergodicity is
restored, and thus the average dimer velocity 
must be zero. But now -- quasi as a precursor of the 
spontaneous deterministic transport -- 
the diffusion coefficient diverges for asymptotically weak noise.
Finally, we will demonstrate that upon application
of a static bias force, 
the system may respond with motion opposite to 
that force, i.e. it exhibits so-called
absolute negative mobility.
The main and indispensable prerequisite for 
all these effects is a non-convex interaction 
potential of the two particles.
This quite unexpected and remarkable condition also seems 
to explain why such effects have not been ``accidentally''
discovered in previous studies of overdamped dimer motion.
An exception is the recent absolute negative mobility effect 
in Ref. \cite{jan11} which  will be discussed in more 
detail in Sec.~X.

\section{One-dimensional Model}
With the exception of Sec.~IX, we will always focus on the 
one-dimensional motion of two coupled, identical particles with 
coordinates $x_i$ ($i=1,2$), modeled by the Langevin equations
\begin{equation}
\eta\, \dot x_i(t) = -\frac{\partial U_{tot}(x_1(t),x_2(t))}{\partial x_i} 
+ F + f(t) + \xi_i(t) \ .
\label{1}
\end{equation}
In doing so, inertia terms $m\ddot x_i(t)$ are considered as negligible
(overdamped limit), $\eta$ is the viscous friction (damping) 
coefficient of the single particles, and $F$ a static 
``bias''-force.
Furthermore, $f(t)$ represents an external driving with 
time period $\tau$, satisfying the symmetry 
condition 
\begin{equation}
f(t+\tau/2)=-f(t) \ ,
\label{1a}
\end{equation}
and thermal fluctuations 
are modeled by independent Gaussian noises $\xi_i(t)$ of
zero mean and respecting the fluctuation-dissipation relation
\begin{equation}
\langle\xi_i(t)\,\xi_j(s)\rangle = 2\, \eta\, k_BT\ \delta_{ij}\,\delta(t-s) \ ,
\label{2}
\end{equation}
where $k_B$ is Boltzmann's constant and $T$ the ambient temperature.
The potential energy $U_{tot}(x_1,x_2)$ is of the form
\begin{equation}
U_{tot}(x_1,x_2)= U(x_1)+U(x_2)+\uint(x_1-x_2)
\label{3}
\end{equation}
with a {\em spatially symmetric} and $L$-periodic 
single-particle potential $U(x)$ and
an interaction potential $\uint (y)$ which will be
further specified later on.

For the sake of convenience, we will often
adopt units of time, length, temperature,
and energy so that
\begin{equation}
\eta=1,\ L=2\pi,\  k_B=1,\ \max_x U(x)-\min_x U(x) = 2 \ .
\label{4}
\end{equation}
When specific examples will be needed, we will 
furthermore focus on
\begin{eqnarray}
U(x) & = & \cos(x)
\label{5}
\\
f(t) & = & A \sin(\omega t)
\label{6}
\end{eqnarray}
with $\omega:=2\pi/\tau$, and either a Lennard-Jones 
or a quartic interaction potential of the form
(see also Fig. 2)
\begin{eqnarray}
\uint  (y) & = & (\lambda/y)^{12} - 2(\lambda/y)^6 \ ,
\label{7}
\\
\uint (y) & = & c\, (y-\Delta)^3 (y-\Delta-2) \ ,
\label{8}
\end{eqnarray}
with positive parameters 
$c$, $\lambda$, $\Delta$.
We expect and explicitly verified in a few cases that
a large variety of similar potentials $U(x)$, $\uint (y)$
and forces $f(t)$ lead to qualitatively unchanged results
(see also Secs.~VI and IX).

\begin{figure}
\epsfxsize=0.95\columnwidth
\epsfbox{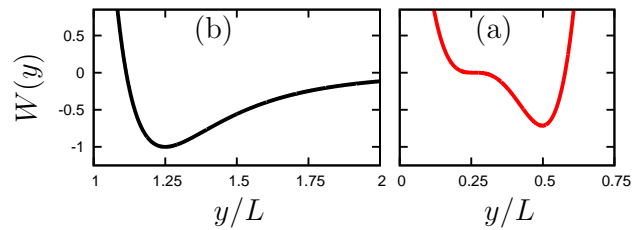} 
\caption{
(Color online)
(a) The Lennard-Jones interaction potential (\ref{7}) for $\lambda=1.25\, L$.
(b) The quartic interaction potential (\ref{8}) for $c=0.428$ and $\Delta=0.26\, L$ (see also Fig. \ref{fig3}). }
\label{fig2}
\end{figure}

The Lennard-Jones potential (\ref{7}) is a well established
prototype model for the interaction of atoms.
In our present context, it has the only, and mainly puristic,
shortcoming that for any finite thermal noise, a dimer
ultimately dissociates after sufficiently long time
into two separate monomers.
Our second example (\ref{8}) is free of this problem
and, moreover, exhibits a slightly richer ``intra-well''
structure of the potential landscape, see Fig. 2.
In particular, it has a metastable inflection point
at $y=\Delta$, thus mediating between a ``purely'' 
monostable and a ``genuine'' bistable interaction 
potential. 
We thus can expect that the results we will find for (\ref{8})
are still qualitatively recovered for suitable monostable 
as well as bistable interaction potentials
(see Sec.~VI.C).

Note that interaction potentials in real systems
are expected to satisfy the symmetry $W(-y)=W(y)$.
In contrast, we also admit non-symmetric $W(y)$,
e.g. in (\ref{8}), for the mere sake of 
convenience later on.
However, when dealing with asymmetric $W(y)$,
we tacitly restrict ourselves to cases
for which a change of sign of $x_1-x_2$ in 
(\ref{3}) is a negligibly rare event in the 
dynamics (\ref{1}), e.g. by admitting only 
sufficiently large $c$-values in (\ref{8}).
In other words, the two monomers do not pass 
each other in practice, and thus only $y$-values of
one sign, say $y>0$, actually matter in $W(y)$.

Later on, our main focus will be on non-convex 
interaction potentials $W(y)$.
Various physical realizations of such potentials 
will be discussed in Sec.~V.
Depending on the particular system, 
the external driving forces $F$ and $f(t)$ may be realized by 
hydrodynamic flows, traveling optical potentials,
electrical fields, magnetic fields,
or mechanical agitation.

\section{Basic notions}
Given the model (\ref{1})-(\ref{3}), it is natural to go
over to center of mass and relative coordinates
\begin{equation}
x:=(x_1+x_2)/2\ ,\ y:=x_1-x_2 \ .
\label{9}
\end{equation}
In the special case (\ref{4})-(\ref{6}), their
equations of motion take the form
\begin{eqnarray}
\dot x(t) & = & \sin(x) \cos(y/2)+A \sin(\omega t)+F+\sqrt{T}\tilde \xi_1(t)
\qquad
\label{10}
\\
\dot y(t) & = & 2 \cos(x) \sin(y/2)- 2\uint '(y) + 2\sqrt{T}\tilde\xi_2(t)
\qquad
\label{11}
\end{eqnarray}
where $\tilde\xi_{1,2}(t)$ are independent, delta-correlated Gaussian noises.

A first main quantity of interest is the time averaged
velocity of the dimer's center of mass,
\begin{equation}
v:=\frac{\tau}{L}\lim_{t\to\infty}\frac{1}{t}\int_0^t dt'\ \dot x(t') \ ,
\label{12}
\end{equation}
expressed as a dimensionless multiple of the spatial and temporal
periods $L$ and $\tau=2\pi/\omega$ for reasons of
convenience later on.

For any finite noise strength $T$,
the dynamics (\ref{1}) is ergodic, i.e. the velocity
(\ref{12}) is (with probability one) equal to the ensemble 
averaged velocity.
In particular, the velocity is independent
of the initial conditions $x_i(0)$ and 
of the realization of the noise $\xi_i(t)$.

In contrast, in the deterministic limit ($T=0$), the velocity (\ref{12})
may -- as we will see later -- still depend on the initial conditions 
$x_i(0)$, implying that ergodicity of the dynamics (\ref{1}) is broken.

A second quantity of interest is 
the diffusion coefficient of the dimer's center of mass,
\begin{equation}
D:=\lim_{t\to\infty}\frac{\langle x^2(t)\rangle
-\langle x(t)\rangle^2}{2t} \ ,
\label{13}
\end{equation}
where $\langle\cdot\rangle$ indicates an ensemble average over
different realizations of the noise $\xi_i(t)$.
For any $T>0$, ergodicity implies that an additional 
averaging over different initial condition $x_i(0)$ 
is possible but actually superfluous in (\ref{13}).

Next, we turn to the notion of spontaneous symmetry breaking.
Generally speaking, this notion refers to a description (model)
of a system in terms of some mathematical equations which exhibit 
a certain symmetry (invariance), while their solutions 
(system states) do not exhibit this symmetry.
The simplest example is the equations 
\begin{equation}
a+b=1\ ,\ a\, b=0
\label{13a}
\end{equation}
which are clearly symmetric in $a$ and $b$, while their solutions,
namely $(a,b)=(1,0)$ and $(a,b)=(0,1)$, both break this symmetry.
It also exemplifies the general rule that different symmetry breaking
solutions are related to each other via the original system symmetry.
The simplest physical example is tossing a coin, illustrating
that the decision about which state 
the system actually selects is usually rooted in the 
initial conditions and/or some ``random effects''.

Returning to our dynamics (\ref{1})-(\ref{3}), we observe 
that it satisfies the following basic symmetry property:
If $x_{1,2}(t)$ is a solution of (\ref{1}) then 
-- due to the symmetry of $f(t)$ in (\ref{1a}) and of 
$U(x)$ mentioned below (\ref{3}) --
also $x^\ast_{1,2}(t):=-x_{2,1}(t+\tau/2)$ will be
a solution of (\ref{1}), provided $F$ is replaced 
by $-F$ and $\xi_{1,2}(t)$ by
$\xi^\ast_{1,2}(t):=-\xi_{2,1}(t+\tau/2)$.

For $T=0$ (deterministic limit) and $F=0$ 
(unbiased system) the dynamics (\ref{1})-(\ref{3})
is thus
-- up to an irrelevant shift of the time-origin
and a relabeling of the (identical) particles --
symmetric under spatial inversion.
It follows that for any possible value of the velocity
$v$ according to (\ref{12}), also the opposite
velocity $v^\ast=-v$ occurs with the same probability
(under the tacit assumption of an analogous
equal a priori likelihood of the initial condition
$x_i(0)$ and $x^\ast_i(0)$).
Since, as mentioned above, different initial
conditions indeed may lead to different 
velocities, it follows that our symmetric system
may give rise to solutions which break the
symmetry of the system by exhibiting a finite 
velocity in either one or the other direction.
As usual (see above), we will refer to such a case as 
{\em spontaneous symmetry breaking} 
and to the concomitant dimer motion as 
{\em spontaneous transport}.

Turning to $T>0$ (and arbitrary $F$), 
the above mentioned fact that the velocity
$v$ from (\ref{12}) is independent of the initial
conditions and the realization of the noise
results in the basic symmetry property
that $F\mapsto -F$ implies $v\mapsto -v$.

In particular, in the unbiased case $F=0$
we recover $v=0$ for any $T>0$.
Regarding finite bias forces $F$, one could
naively expect that 
the velocity (\ref{12}) now should take on a finite value 
and that its sign should always correspond to that of $F$.
The quite unexpected contrary behavior, namely
a velocity $v$ opposite to the applied force,
is termed {\em absolute negative mobility}
\cite{eic05}.

The main aim of our paper is to demonstrate that the
dynamics (\ref{1}) indeed may give rise to
spontaneous transport and to absolute
negative mobility, and that there exists in fact
a close connection between these two quite 
astonishing effects.
A comparison with related previous works 
is provided in Sec.~X.

\section{No-go theorems for convex interaction potentials}
In this section we demonstrate that spontaneous transport and absolute
negative mobility (as specified in the previous section) are
ruled out for any dynamics (\ref{1})-(\ref{3}) with a convex 
interaction potential $W(y)$, i.e.
\begin{equation}
W''(y)>0\ \mbox{for all $y$.}
\label{14}
\end{equation}

In order to exclude spontaneous transport, it is sufficient to demonstrate
that the velocity (\ref{12}) in the deterministic limit ($T=0$) is
unique (independent of the initial conditions).
Proceeding by way of a proof by contradiction, we assume that there exist 
two solutions of (\ref{1})-(\ref{3}), called $x_i(t)$ and $x'_i(t)$, with
different velocities $v$ and $v'$ according to (\ref{12}), say
$v'>v$ without loss of generality.
Since for any integer $n$, also
$x_i(t)+nL$ solves (\ref{1})-(\ref{3}) and yields the same velocity
$v$ in (\ref{12}), we may without loss of generality assume that
the initial conditions satisfy 
$x_1(0),\, x_2(0) > x'_1(0),\, x'_2(0)$.
Since $v'>v$, the particles $x'_i(t)$ must ``catch up'' with the particles
$x_i(t)$ in the course of time.
Without loss of generality, the indices $i$ can be chosen 
so that it is $x'_1(t)$ which catches up with $x_1(t)$ for the first time
at some $t=t^\ast$, i.e. 
\begin{eqnarray}
x'_1(t^\ast) & = & x_1(t^\ast)
\label{15}
\\
x'_2(t^\ast) & \leq & x_2(t^\ast)
\label{16}
\\
\dot x'_1(t^\ast) & \geq & \dot x_1(t^\ast) \ .
\label{17}
\end{eqnarray}
The equality sign in (\ref{16}) can be excluded, since otherwise
(\ref{15}) and the dynamics (\ref{1})-(\ref{3}) (with $T=0$)
would imply that $x'_i(t)=x_i(t)$ for all $t$, and thus $v'=v$.
With $y:=x_1(t^\ast)-x_2(t^\ast)$ and $y':=x'_1(t^\ast)-x'_2(t^\ast)$ it follows that
\begin{equation}
y'>y \ ,
\label{18}
\end{equation}
and by introducing (\ref{15}) into
(\ref{1})-(\ref{3}) (with $T=0$) we find
\begin{eqnarray}
\dot x'_1(t^\ast) -  \dot x_1(t^\ast) & = & -W'(y')  + W'(y) \ .
\label{19}
\end{eqnarray}
In view of (\ref{18}) and (\ref{14}) we thus can infer
that $\dot x'_1(t^\ast) - \dot x_1(t^\ast) < 0$, 
in contradiction to (\ref{17}).
This concludes our proof that spontaneous 
transport is excluded.

Next we analogously proof by contradiction that (\ref{14})
rules out absolute negative mobility for the dynamics
(\ref{1})-(\ref{3}) with $T>0$.
We thus consider a solution $x_i(t)$
of (\ref{1})-(\ref{3}) with velocity $v$ according to (\ref{12})
and a second solution $x'_i(t)$ 
with $F'$ instead of $F$ in (\ref{1}) and a 
resulting velocity $v'$.
Assuming absolute negative mobility, it follows that
two such solutions must exists with the property 
that $F'<F$ and $v'>v$.
In particular, we can assume the same realization of the noise
$\xi_i(t)$ in (\ref{1}) for $x_i(t)$ and for $x_i'(t)$,
since -- as mentioned in Sec.~III -- the velocity 
(\ref{12}) is independent of the realization of 
the noise.
As before, $x_1(0),\, x_2(0) > x'_1(0),\, x'_2(0)$ can be 
taken for granted without loss of generality and $v'>v$ then
implies the existence of a time point $t^\ast$ with the properties
(\ref{15})-(\ref{17}).
With $y:=x_1(t^\ast)-x_2(t^\ast)$ and $y':=x'_1(t^\ast)-x'_2(t^\ast)$
it follows that
\begin{equation}
y'\geq y \, ,
\label{20}
\end{equation}
but, unlike before, $y'=y$ cannot be excluded 
now. From (\ref{1})-(\ref{3}) we can infer that
\begin{eqnarray}
\dot x'_1(t^\ast) -  \dot x_1(t^\ast) & = & -W'(y')  + W'(y) + F' - F \ .
\label{21}
\end{eqnarray}
Combining (\ref{14}) with (\ref{20}) yields $-W'(y')  + W'(y)\leq 0$
and with $F'<F$ it follows that $\dot x'_1(t^\ast) -  \dot x_1(t^\ast)<0$,
in contradiction to (\ref{17}).
We thus have proven that absolute negative mobility is impossible.

We remark that the above arguments can be readily generalized 
to more than two and/or non-identical particles with convex
nearest-neighbor interaction potentials.
In particular, in the context of the Frenkel-Kontorova model, the 
so-called Middelton's no passing rule has been proven along 
similar lines, 
see \cite{middleton91,middleton92,middleton93,floria96,baesens98}
and further references therein.

\section{Non-convex interaction potentials}
Non-convex interaction potentials occur naturally 
in various situations:
First, this is the case for many common models of
basic molecular interactions 
\cite{qin00,boisvert97,pijper07}, 
such as Lennard-Jones 
\cite{lj} (cf. (\ref{7})) or Morse potentials
\cite{pijper05,craig06}.
Second, less basic but still bistable molecular interaction potentials 
arise, e.g., in the modeling of random walkers in motor proteins
\cite{ajd94,pes95,derenyi96,julicher97,mog98,ast99,str99,li00,els00,klumpp01,sym-mm}, 
and have also been considered in dimer ratchet models
\cite{mateos04,menche06,retkute06,pototsky09}.
Third, more complex dimers may have several stable configurations,
or the dissociation process may involve additional (fast) internal 
degrees of freedom that lead to a non-convex effective interaction 
potential, as considered theoretically e.g. in \cite{denisov02}.
Experimentally, e.g. in \cite{san99} a water droplet was positioned 
on a periodic surface and its shape (internal degree of freedom) was 
periodically driven by externally applied electric fields.
Lastly, a spatially one dimensional model
may serve as an approximation
for the full three-dimensional dynamics
of the dimer. The main effects of
the other degrees of freedom are incorporated
into the ``effective'' interaction forces
between the monomers, resulting in
a more complex, non-convex structure of the potential (cf.\ Sec.~IX).

\section{Spontaneous symmetry breaking}
We consider the dynamics (\ref{1})-(\ref{3}) without 
bias ($F=0$) in the deterministic limit ($T=0$).
We furthermore focus on non-convex interaction 
potentials $W(y)$, for instance (\ref{7}), (\ref{8}).
Is it possible that a solution $x_i(t)$ spontaneously
breaks the spatial symmetry of the system and exhibits 
a finite velocity $v$ according to Eq.~(\ref{12})?
While there seems to be no a priory reasons why not,
the actual existence of such solutions appears to 
us quite astonishing nevertheless.
In accordance with this intuitive expectation, their 
existence turns out to be indeed restricted
to relatively small parameter regions in the deep 
non-linear regime, 
thus making any further progress by analytical 
means quite hopeless.
Therefore, we focus on the discussion of our
extensive numerical results in the first place
and only a posteriori provide a simplified 
intuitive picture of what is going on.

\subsection{Numerical example}
Figure 3 presents typical results for the quartic
potential from (\ref{8}) (see also Fig.\ 2b). 
For each depicted point (``pixel'') in Fig.\ 3,
we have calculated the velocity according to 
(\ref{12}) for 20 trajectories of length $500\,\tau$ 
with randomly chosen initial conditions.
To eliminate ``transient'' effects,
we furthermore excluded an initial
time-interval of length $150\,\tau$ for
each trajectory contained in the average
in (\ref{12}).
The black ``islands''
indicate parameter regions where all initial conditions 
which we numerically tried out 
ended up on periodic attractors
exhibiting spontaneous symmetry breaking, 
i.e.\ a finite net velocity of the rational form
$v=\pm n/m$ with positive integers $n$, $m$
in the specific dimensionless units 
from (\ref{12}).
We note that spontaneous symmetry breaking
may occur also with other types of attractors,
e.g.\ chaotic attractors which emerge from the periodic ones by
period doubling cascades \cite{strogatz} and thus
still carry a rational net velocity of the form
$v=\pm n/m$. However, they are not visible in Fig.\ 3,
as their parameter regions are extremely small.
Furthermore, we found that the black islands 
in Fig.\ 3 are largely dominated by
the velocities $v=\pm 1$.
The light grey ``sea'' around the black islands
indicates parameter regions without 
spontaneous symmetry breaking, 
i.e.\ all probed initial conditions
exhibited a net velocity of $v=0$
(in fact, every trajectory $x(t)$ remained
bounded, and, in particular, 
did not exhibit deterministic diffusion).
The quite small remaining regions in mid-dark 
grey indicate coexistence of solutions with
$v=\pm n/m$
(like in the black islands) and
with $v=0$ (like in the light grey sea).
The tiny (almost invisible) white regions and single 
pixels in Fig.\ 3 indicate that
the dimer trajectory was neither
bounded nor did it reach
a periodic attractor
within the maximal numerical simulation time 
of $1000\,\tau$.
Since those parameter regions are so small, 
we will not consider them any further.

\begin{figure}
\epsfxsize=0.95\columnwidth
\epsfbox{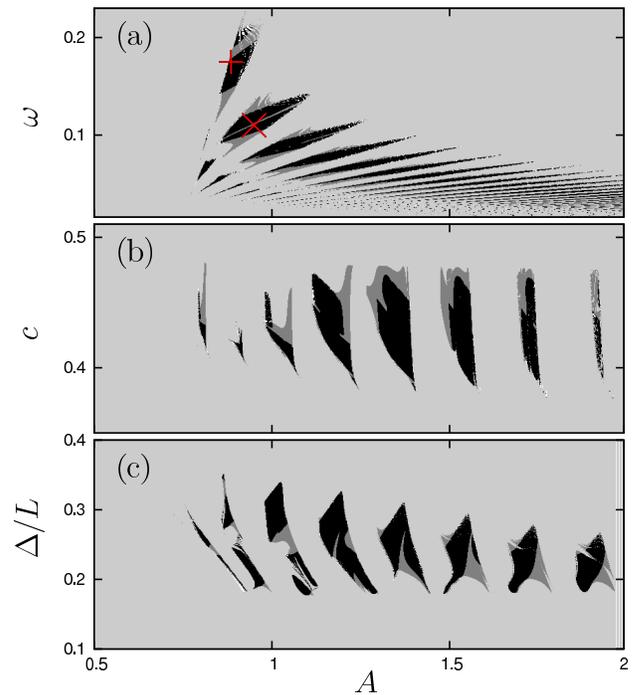} 
\caption{(Color online)
Velocity (\ref{12}) obtained by numerically solving
(\ref{10}), (\ref{11}) with $F=0$, $T=0$, and (\ref{8})
for various parameter values.
(a) Variable driving amplitude $A$ and
frequency $\omega$ with fixed parametes
$c=0.428$, $\Delta=0.26L$
(these ``odd'' values are chosen on purpose
to avoid all ``accidental symmetries'' or 
other sorts of ``special cases'').
The two crosses ``$+$'' and ``$\times$'' indicate the specific parameter 
values adopted later in Fig.~4 and Fig.~6, respectively.
(b) Variable driving amplitude $A$ and
interaction potential parameter $c$ with
fixed parameters $\omega=0.064$, $\Delta=0.26\, L$.
(c) Variable driving amplitude $A$ and
interaction potential parameter $\Delta$ with
fixed parameters $\omega=0.064$, $c=0.428$.
Black: Finite velocities $v=\pm n/m$
with positive integers $n$ and $m$.
Light grey: Velocities $v=0$.
Intermediate grey: Coexistence of finite
velocities $v=\pm n/m$ and $v=0$.
For further details see main text.
}
\label{fig3}
\end{figure}

The negative parameter regimes (omitted in Fig.\ 3) 
are either unphysical (e.g.\ $c$ in (\ref{8})), follow 
by symmetry (e.g.\ $A$ and $\omega$ in (\ref{10})),
or by periodicity (e.g.\ $\Delta$ in (\ref{8})).
For all the remaining (non-negative) parameters outside the 
range plotted in Fig.\ 3, either only solutions with $v=0$ were 
obtained, or -- in the case of large $A$-values --
the continuation is qualitatively quite obvious.
By changing the values of those parameters which are
held fixed in each of the three plots in Fig.\ 3, 
also the shape and positions of the ``black islands'' 
change considerably.
E.g., we have found that
for every given $0.001<\omega<2$ 
it is possible to
adapt the remaining parameters so that spontaneous
symmetry breaking is realized
(even smaller $\omega$-values become numerically very 
expensive and have therefore not been investigated
any further). From 
this viewpoint, the range of symmetry breaking 
parameters is actually not so small as a first glance
on Fig.\ 3 may suggest.

\subsection{Basic mechanism}
In order to illustrate the characteristic features of the dimer motion
when the symmetry
is spontaneously broken, we
focus, as a typical example, on the parameter values
of the driving indicated by
the ``$+$'' symbol in Fig.\ 3a 
and on the specific interaction potential $W(y)$
depicted in Fig.\ 2b.
For these parameter values,
two periodic attractors with
net velocity $v=\pm 1$ coexist, which
break the symmetry
of the underlying equations of motion
(\ref{10}), (\ref{11}).
Figure 4 collects 9
snapshots of the associated dynamical evolution
of the dimer during one driving period $\tau$
on the attractor with $v=+1$.
At the beginning of the cycle,
the dimer extension is close to its equilibrium
value of $y = L/2=\pi$ (see also Fig.\ 2b and Eq. (\ref{4})).
In this configuration, the forces due
to the external periodic potential acting on the
two particles are almost identical in modulus but
have opposite signs, so that the center of mass
motion of the dimer is practically not affected
by the periodic potential
(cf.\ Eq.\ (\ref{10}) with $y\approx L/2=\pi$),
whereas its impact on the relative
coordinate $y$ is maximal (cf.\ Eq.\ (\ref{11})).
As a consequence,
the dimer as a whole essentially follows the time-dependent
driving force, while it is slightly stretched and compressed
by the periodic potential landscape.
This type of behavior can be observed in Fig.\ 4 during the
first half of the driving period $\tau$ and
results in a displacement of the dimer's center by a
little more than one spatial
period in the positive $x$ direction.
The deviations from the equilibrium length
$y = L/2$ are stabilized very quickly due to the 
convex character of the
interaction potential $W(y)$ for $y\approx L/2$.
However, the non-convexity of $W(y)$ for
$y$-values around $y=\Delta=0.26L$ allows
the dimer to be eventually compressed to about
$L/4$ in length,
and to maintain
this ``metastable'' state for almost the complete
second half of the driving period.
During this phase, the dimer is trapped motionless in a
minimum of the rocking potential $U(x) - xf(t)$.
At about $6\tau/8$, when the left potential barrier of the minimum is nearly
balanced by the external driving (although the minimum
actually never disappears because the driving amplitude $A$
is smaller than unity), the dimer starts to relaxate back to
its equilibrium length $y = L/2$ by way of ``pushing the left monomer 
over the barrier''. In this manner, the dimer reaches its 
initial configuration, but with the center being shifted
by $n=1$ spatial period to the right
after $m=1$ cycle of the driving,
resulting in $v=+1$.

\begin{figure}
\epsfxsize=0.95\columnwidth
\epsfbox{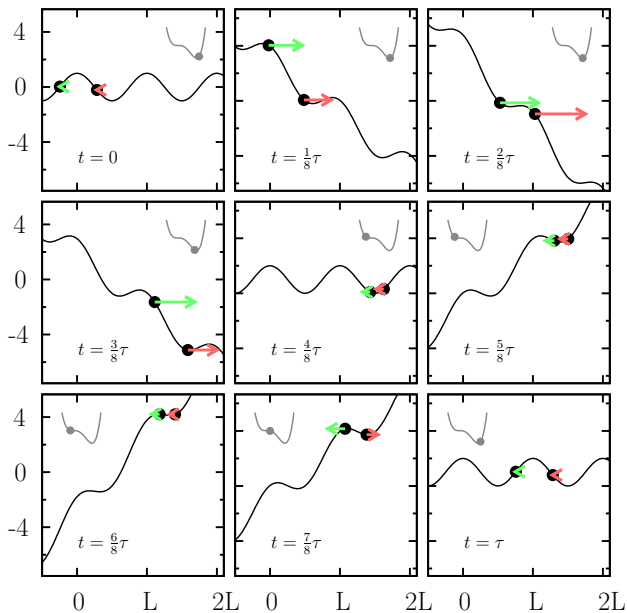} 
\caption{(Color online)
Time evolution of a typical 
spontaneous symmetry breaking periodic attractor of 
(\ref{10}), (\ref{11}) with interaction potential 
(\ref{8}) and the specific parameter values indicated 
by the ``$+$'' symbol in Fig.~3a, 
i.e. $\omega=0.175$, $A=0.885$ 
and all remaining parameters as in Fig. 3a.
The black dots indicate $x_1(t)$ (right dot) and $x_2(t)$ (left dot),
the arrows $\dot{x_1}(t)$ and $\dot{x_2}(t)$, and
the black lines the tilted periodic potential 
$U(x)-(x-L/2)f(t)$.
The dark grey ``insets'' indicate the interaction 
potential (\ref{8}) (grey line) along with the 
instantaneous dimer length $y(t)$ (grey dot).
}
\label{fig4}
\end{figure}

From these observations we get an intuitive idea,
in addition to the formal arguments of Sec.~IV, of why
non-convexities in the interaction potential are necessary
for spontaneous symmetry breaking. They promote the existence
of a ``metastable'' dimer state with dynamical behavior
different from that of the equilibrium configuration.
The dimer motion in the periodic potential landscape under
the action of the external driving induces
periodic transitions between that ``metastable'' state
and the equilibrium configuration, associated with
distinct phases in the dimer motion and thus resulting
in an overall symmetry breaking behavior.
The above example has been chosen to most clearly
illustrate this mechanism, as the specific parameter
values give rise to an equilibrium configuration
with practically ``free'' motion of the dimer's center,
whereas in its ``metastable'' state, the dimer is
trapped motionless by the potential landscape.
Such a clear-cut distinction between the two states
is, however, not necessarily required.
Figure 3 shows that spontaneous symmetry breaking
occurs in quite large parameter regions,
where the
``metastable'' state is ``sufficiently different''
from the equilibrium configuration of the dimer.

From our discussion so far, it is evident
that average dimer velocities larger in modulus than $|v|=1$
should be observable. For instance, for slower driving
(smaller $\omega$) the dimer may be able
to travel several spatial periods during the
first half of the driving cycle, while it gets trapped
during the second half of the cycle. Indeed,
dimer velocities with $v=\pm 2$ are present
in the black regions of Fig.\ 3a for small $\omega$ and small $A$ values
(in Fig.\ 3, the information on the actual value of non-vanishing
average velocities in the black regions is not included though).
Likewise, rational velocities with $|v|<1$
can, e.g., be found for fast driving at the edges of the black regions
towards large $\omega$ and $A$ values.

While these considerations are helpful to gain an
intuitive picture of what is going on, they obviously
do not prove that a specific symmetry breaking solution
is dynamically stable (an attractor) and that, besides
its twin brother with opposite velocity, no other stable
solutions exist. Any such ``more rigorous'' reasoning would 
have to depend on the precise quantitative values of all
model parameters and thus seems almost impossible.

\subsection{Generalizations}
We finally point out that the specific shape of the interaction potential
is not crucial for the existence of spontaneous symmetry 
breaking, as long as it is non-convex.
For instance, we have found
spontaneous symmetry breaking attractors for various 
modifications of (\ref{8}), including
the obvious situation
where it has a second minimum,
and also the ``extreme''
case of an ``almost convex'' potential
with a curvature that is positive everywhere
except near the equilibrium length $y\approx L/2$.
Moreover, spontaneous symmetry breaking is also
supported by completely different non-convex
interaction potentials.

\begin{figure}
\epsfxsize=0.95\columnwidth
\epsfbox{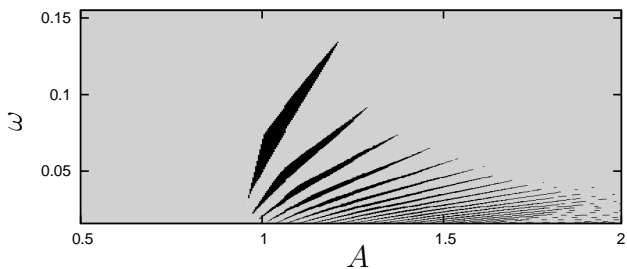} 
\caption{Same as Fig. 3a but for the Lennard-Jones 
interaction potential (\ref{7}) with $\lambda=1.25\, L$.
Only attractors that can be reached by dimers initially close
to their equilibrium lengths are shown, see text.
}
\label{fig5}
\end{figure}

As a specific example, we
now turn to our numerical results for
the Lennard-Jones potential (\ref{7}), see also Fig.~2a.
For large distances $y=x_1-x_2$ of the monomers,
their interaction becomes negligibly small.
As expected, we found numerically that in this case 
each monomer exhibits a periodic back-and-forth 
motion in the long time limit and thus the 
velocity $v$ from (\ref{12}) vanishes.
Since these solutions are of little interest,
we henceforth use only initial conditions such that
the dimer length is initially close to the 
equilibrium length $\lambda$ of the potential (\ref{7}).
More precisely, we only considered 
$y(0)\in [0.9\cdot\lambda,1.3\cdot\lambda]$
and observed that the distance
$y(t)$ always remained reasonably small to
consider the dimer as ``not dissociated''
(a rigorous dissociation criterion for the
potential (\ref{7}) seems hard to define).
The so obtained numerical results are qualitatively
very similar to those for the quartic potential (\ref{8}),
although the quantitative details are different.
As an example, Fig.~5 depicts our findings
for the Lennard-Jones potential (\ref{7}) corresponding 
to those from Fig.~3a for the quartic potential (\ref{8}).
In particular, we recover spontaneous symmetry breaking
within a quite notable parameter range.
The first main differences compared to the quartic potential
is that we did not find coexistence of solutions with
$v=0$ and $v\neq 0$.
The second main difference is that all the spontaneous 
symmetry breaking solutions that we found exhibited 
the specific velocities $v=\pm 1$ for all the parameters 
that we tested.

\section{Giant diffusion}
We still focus on the unbiased ($F=0$) dynamics (\ref{1})-(\ref{3}), 
but now consider finite noise strengths ($T>0$),
implying $v=0$ for all model parameters and
all initial conditions (see Sec.~III).
The reason is that any attractor of the 
deterministic dynamics -- in particular those
giving rise to spontaneous symmetry breaking --
now becomes metastable due to noise induced
transitions between them.
While the velocity (\ref{12}) is thus trivial, 
the quantity of foremost interest is now the 
diffusion constant (\ref{13}).

For large noise ($T\gg 1$) in (\ref{1})-(\ref{4}),
the effects of the ``substrate potential''
$U(x)$ become negligible, hence the diffusion
asymptotically approaches the value for the
``free'' thermal diffusion of the dimer,
namely $D=k_B T/2\eta$.

Much more interesting is the opposite asymptotics
of small $T$, specifically in the case of spontaneously 
broken symmetry without coexisting deterministic
solutions with $v=0$, as exemplified by the
black islands in Figs. 3 and 5.
Focusing on this case, a solution $x_i(t)$ of the
noisy dynamics (\ref{1})-(\ref{3}) closely
follows most of the time either a deterministic
solution exhibiting $v>0$ or its twin brother 
with $v<0$, switching between them every now 
and then by way of noise activated transitions.
For sufficiently weak noise, these transitions become
very rare events and are captured very well by
a rate process with a certain transition rate $k$. 
A similar calculation as 
in Ref. \cite{kap82} then yields for the 
diffusion coefficient (\ref{13}) the result
\begin{equation}
D=\frac{L^2}{2 \tau^2\, k} \ .
\label{22}
\end{equation}
The transition rate itself is of the usual
Arrhenius-form
\begin{equation}
k=\nu_0\ \exp\{-\Delta\phi/T\}
\label{23}
\end{equation}
with a $T$-independent generalized activation
energy $\Delta\phi$ 
\cite{lehmann02}
and an ``attempt frequency'' $\nu_0$ with a
rather weak $T$-dependence (at most algebraic).
Combining (\ref{22}) and (\ref{23}), we thus
arrive at the analytical prediction 
\cite{spe12}
of an extremely strongly diverging diffusion coefficient
for asymptotically weak noise strengths $T$
(essential singularity).
The numerical findings in Fig.~6 confirm this
prediction very well.

\begin{figure}
\epsfxsize=0.95\columnwidth
\epsfbox{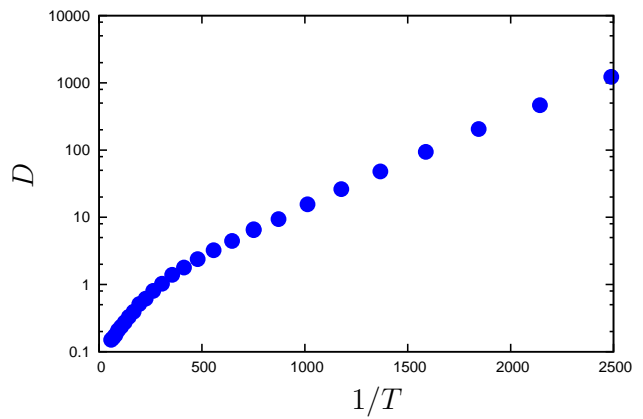} 
\caption{(Color online)
Arrhenius plot (logarithm of $D$  versus $1/T$) 
of the diffusion coefficient (\ref{13}) by numerically
solving the unbiased ($F=0$) dynamics (\ref{10}), (\ref{11}) 
with quartic interaction potential (\ref{8}) and the
specific parameter values indicated by the ``$\times$'' symbol
in Fig. 3a,
i.e. $\omega=0.11$, $a=0.95$ 
and all remaining parameters as in Fig. 3a.
}
\label{fig6}
\end{figure}

\section{Absolute negative mobility}
As before, we consider the dynamics 
(\ref{1})-(\ref{3}) with $T>0$,
but now also in the presence of a 
non-vanishing static bias $F$.
For symmetry reasons (see Sec.~III) we 
henceforth focus on $F>0$.
Our main question is: how does the
unperturbed velocity ($v=0$ for $F=0$) 
respond
to a perturbation $F>0$, and, in particular, 
is it possible to obtain absolute
negative mobility, i.e. $v<0$ ?

To be specific, we revisit the deterministic
($T=0$) and unbiased ($F=0$) dynamics (\ref{10}), (\ref{11})
for the quartic potential (\ref{8}), resulting
in the velocity diagram from Fig.~3a.
We recall that every black island in this figure
indicates the coexistence of two symmetric deterministic 
attractors, one with $v>0$ and the other with $v<0$,
and that all solutions (for arbitrary initial 
conditions) approach one of these two attractors 
in the long time limit, i.e. all solutions
exhibit either $v>0$ or $v<0$.
As soon as $F>0$, the symmetry of the system is 
broken and hence the degeneracy of the $v\neq 0$ 
solutions lifted.
Accordingly, by gradually increasing $F$, the ``island''
in parameter space describing solutions with $v>0$
will start to differ more and more from the $v<0$ ``island''.
Every subset of the latter island with $v<0$ which is
located ``outside'' of the $v>0$ island is clearly a 
quite promising candidate for absolute negative mobility 
at finite (sufficiently small) $T$.
Fig.~7a confirms that such subsets indeed exist 
and Figs.~7b,c show that they indeed give rise to absolute 
negative mobility.

\begin{figure}
\epsfxsize=0.95\columnwidth
\epsfbox{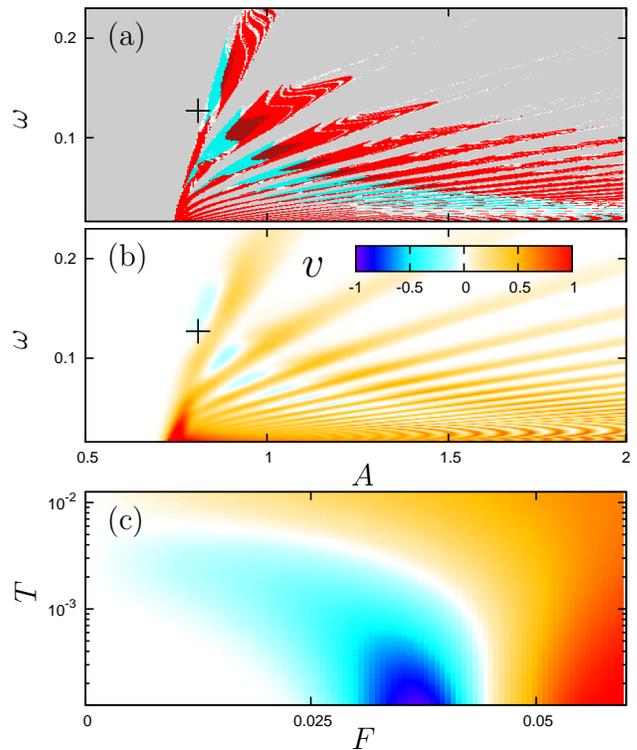} 
\caption{(Color) (a): Same as Fig.~3a, but for a finite bias
$F=0.01$. 
Red: Velocities $v>0$. {\Cyan}: Velocities $v<0$.
Red and {\cyan} overlapping: Coexistence
of solutions with opposite average velocities.
Grey: Velocities $v=0$.
Coexistence of $v\neq 0$ and $v=0$ is resticted to 
negligibly small parameter regions and is not shown in 
this figure.
(b): Same as in (a) but for a finite noise strength 
$T=0.005$. 
The value of the velocity $v$ from (\ref{12}) is 
independent of the initial conditions and indicated 
by the color code. In particular, blue regions 
correspond to absolute negative mobility 
($v<0$ for $F>0$).
(c) Dependence of the velocity $v$ on the bias $F$ and
the noise strength $T$ for fixed driving parameters as
indiciated by the black ``$+$'' in (a) and (b).
}
\label{fig7}
\end{figure}

By comparing Fig.~7a with Fig.~3a we see that
the degeneracy of the $v>0$ and $v<0$ solutions is lifted
by displacements of the red $v>0$ ``islands'' toward
larger $A$ and $\omega$ values and of the cyan $v<0$ 
``islands'' toward smaller $A$ and $\omega$ values.
While the impact of a change in the driving amplitude $A$ on a specific
solution and its stability is hard to grasp intuitively
\cite{spe11},
the existence of a ``purely {\cyan} stripe'' 
outside the border of any ``red island''
toward smaller frequencies $\omega$
can be understood as follows:
In the unbiased case ($F=0$) from Fig.~3a, the lower border of an
island indicates that solutions with $v\neq 0$ will cease 
to exist (become unstable) upon further decreasing the 
driving frequency $\omega$. 
Roughly speaking, the driving thus becomes ``too slow''
compared to the ``eigenspeed'' of the solutions, 
i.e. they are unable to synchronize with the imposed temporal 
periodicity of the driving.
Moreover, if a small bias $F>0$ is switched on, it is quite 
plausible to expect that the  intrinsic ``eigenspeed'' will 
decrease for solutions traveling ``uphill'' (against the force $F$), 
and increase for those traveling downhill. 
As a consequence, at the lower border of an island in Fig.~3a,
those solutions with $v<0$ should survive and those with 
$v>0$ disappear when a small $F>0$ is switched on,
in agreement with Fig.~7a.

Upon further increasing $F$, one expects that the {\cyan} and red 
islands in Fig.~7a change not only their positions, but also their
shapes and in particular their areas, and that for sufficiently
large $F$, the {\cyan} islands will completely disappear.
Likewise, with increasing noise strength $T$, one expects that
the original deterministic solutions become less and less 
relevant.
Indeed, for $T\to\infty$ the effects of the substrate potential 
$U(x)$ in (\ref{3}) become negligible, yielding for the
velocity (\ref{12}) the trivial result $v=F/\eta$.
These quite plausible qualitative features are indeed confirmed 
by our numerical results for a representative example in Fig.~7c.

Turning to the Lennard-Jones potential (\ref{7}), it is quite
plausible and can be confirmed by more detailed calculations
that any solution of (\ref{1}) with $T>0$ results in a 
dissociation of the dimer in the long run, and that this
implies $v>0$ for all $F>0$.
However, for small $T$ and $F$, the lifetime of the dimer
is still very large, and restricting the time-average 
of the velocity in (\ref{12}) accordingly, the same
qualitative features as in Fig.~7 are recovered
(not shown).
Fig.~8 exemplifies a typical solution exhibiting
absolute negative mobility during the extremely
long initial time-period before the dimer dissociates.
The very long life-time of the dimer can be
understood intuitively as follows:
As a relict of the underlying
(dynamically stable) symmetry-breaking attractor, 
the dimer length never increases notably beyond $2$
during the motion shown in Fig.~8.
The binding energy of the dimer at that length is
still $\mathcal{O}(10^{-1})$ (see Fig.~2b)
and thus much larger than the thermal energy corresponding to
$T=0.001$, so that
noise-induced dissociation events are extremely rare.

\begin{figure}
\epsfxsize=0.95\columnwidth
\epsfbox{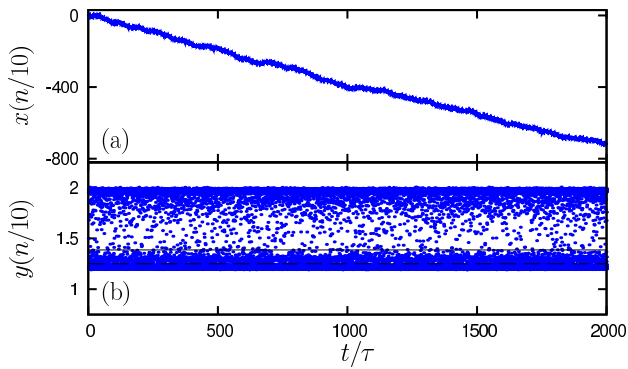} 
\caption{(Color online)
(a) Center of mass $x(t)$ from (\ref{9})
for the dynamics (\ref{10}), (\ref{11}) with
Lennard-Jones interaction potential (\ref{8}).
Remaining parameters:
$A=1$, $\omega=0.06$, $T=0.001$, $F=0.01$.
(b) Dimer length $y(t)$ at the
time points $t=n/10$ for $n=0,1,...,20000$.
}
\label{fig8}
\end{figure}

\section{Outlook on higher dimensional Models}
In this section we address the motion of two coupled 
particles in 2- or 3-dimensional periodic potentials,
e.g. on a crystal surface, in a lattice of optical or 
magnetic traps etc., see also Sec.~I.
A particularly interesting special case arises for
systems with a pronounced anisotropy, causing a strong
preferential direction (``easy axis'') for the
dimer motion and orientation, as exemplified with Fig.~1.

Focusing on this case, the one-dimensional model (\ref{1})-(\ref{3}) 
still describes such systems reasonably well, especially
with the following modification:
The main effect of the omitted additional dimensions
is that the dimer may ``turn around'' by 180$^\circ$, 
albeit with a very low probability per time unit
(due to the strong anisotropy).
This feature can be effectively incorporated into the model 
(\ref{1})-(\ref{3}) by choosing an interaction potential
$W(y)$ which exhibits two minima, one at a 
positive and one at a negative $y$-value, separated by a ``high'' 
energy barrier at $y=0$ (to guarantee ``rare'' transitions).
In other words, a sign change of $x_1-x_2=y$ 
(cf. Eq. (\ref{9})) does not mean that the two particle pass 
``through'' each other, but rather that they perform a 
rotation.
In contrast to our convention at the end of Sec.~II,
we furthermore restrict ourselves to  ``physically realistic'' 
interactions, depending only on the absolute particle 
distance and thus satisfying
\begin{equation}
W(-y)=W(y) \ .
\label{26}
\end{equation}
In other words, $W(y)$ exhibits a symmetric double well
structure and thus is always a non-convex potential.

Numerically, we found that effectively one-dimensional 
models of the above type reproduce all the main effects
from Secs.~VI-VIII.
Essentially, the reason is that once the solutions
of a model as specified in Sec.~II are known, the solutions
of the same model but with $W(y)\mapsto W(y+L)$ follow
immediately and the resulting velocities and diffusion 
coefficients remain unchanged.
Hence, our present double well potential $W(y)$ with
a barrier at $y=0$ can be readily shifted by any
integer multiple of $L$ ``to the right'' until it becomes
similar to the one of the previously considered interaction
potentials (see Sec.~VI.C).

More generally, also dimers consisting of non-identical 
particles can be approximated by an effectively one-dimensional 
model analogous to (\ref{1})-(\ref{3}), except that the friction
coefficients $\eta$ in (\ref{1}) may now be different for 
the two particles, and similarly for the ``coupling strengths''
to the forces $F$, $f(t)$ and the substrate potentials $U(x)$.
As a consequence, one can show that the system satisfies only
a weaker symmetry property than the one from Sec.~III.
Accordingly, a spontaneous breaking of this symmetry
is in many cases rather trivial: Namely,
the dimer exhibits a standard ratchet effect 
since the two particles are different and never
actually exchange their positions during their 
time-evolution \cite{cil01,gehlen08,gehlen09}.
Considerably less trivial at first glance
is the occurrence of giant diffusion and 
absolute negative mobility in such systems.
But given our previous discussions in Secs.~VII and VIII,
these effects are, of course, immediate consequences 
of the spontaneous symmetry breaking.

\section{Conclusions}
We considered a very simple ``minimal'' model
arising in a large variety of different situations
(see Sec.~I):
Two coupled overdamped Browian particles
moving along a periodic substrate potential 
in one or more dimensions.
In spite of the fact that the particles are 
identical and the substrate spatially symmetric,
a sinusoidal external driving may lead to spontaneous 
symmetry breaking in the form of a permanent 
directed motion of the dimer 
(spontaneous transport, see Sec. V)
in the deterministic limit ($T=0$).
Thermal noise ($T>0$) restores ergodicity and thus
zero net velocity, but entails arbitrarily fast 
diffusion of the dimer for asymptotically small
temperatures $T$ (Sec. VII)
Moreover, for small but finite $T$ the dimer responds 
to a static bias force with motion opposite to that 
force (absolute negative mobility, see Sec. VIII).
As proven in Sec. IV, the key requirement for all 
these effects is a non-convex interaction potential 
of the two particles (Sec. V).

Directed transport induced by spontaneous symmetry breaking
as a collective effect of infinitely many 
particles far from equilibrium has been
observed before, e.g., in Refs.\ 
\cite{juelicher95,julicher97,badoual02,reimann99}.
These systems and also the basic physical mechanism are
quite different from our present case (Sec. VI.B).
Spontaneous symmetry breaking induced transport for
periodically driven, underdamped (finite inertia)
systems 
\cite{barone82,kautz96,speer07,nagel08}
is, mathematically speaking, closer related to our 
case (both types of dynamics are two-dimensional, dissipative, 
and non-autonomous), while the underlying physical 
systems (Sec. I) are entirely different.

Diffusion enhancement phenomena are well established already
for a single overdamped particle under the influence of 
an unbiased, temporally periodic driving force 
\cite{gang96,schreier98,reguera02}.
But a diverging diffusion coefficient 
for asymptotically weak noise has been
reported for the first time only very 
recently in \cite{spe12}.
In particular, we remark that,
unlike in \cite{gang96,schreier98,reguera02},
in our present case, the parameter space of diffusion enhancement 
does not approach a set of measure zero
for asymptotically weak noise.

Absolute negative mobility
has been reported in a considerable variety of
experimental systems and theoretical models by now, 
see \cite{eic05} for a recent review.
Within the realm of classical models (no quantum effects),
it has been observed for the first time as a collective 
effect of infinitely many interacting particles 
under far from equilibrium conditions
\cite{collective}.
Later, this model was successfully reduced down to
3 interacting particles \cite{vandenBroeck02}.
Although this model is very different from ours,
the present results for two interacting particles 
may be seen as the next and ultimate reduction step,
since, as demonstrated in \cite{speer07}, 
a single overdamped 
particle in one dimension cannot exhibit 
absolute negative mobility.

A somewhat similar stochastic dynamics of two coupled
Josephson junctions exhibiting absolute negative mobility
has recently been studied in Ref. \cite{jan11}.
This model is mathematically equivalent to the overdamped
dynamics of two coupled particles. However, it exhibits
certain features which are very 
different from our present model:
(i) The two particles are not identical in all respects;
(ii) The interaction forces cannot be derived 
from an interaction potential. Moreover, they 
are spatially periodic, i.e. not approaching zero 
for large particle distances.
Besides those differences in the considered model, 
also the basic mechanism generating absolute 
negative mobility is very different.

Further, mathematically somewhat similar but physically
completely different systems exhibiting absolute negative 
mobility are (i) a single
underdamped (finite inertia) particle in one dimension
\cite{speer07,nagel08,kostur},
(ii) an overdamped particle on a one-dimensional ``meandering'' state space
\cite{eic05,meander} 
(iii) an overdamped particle in 2 
dimensions \cite{ANM2D},
and (iv) a single particle in one dimension with
some kind of ``internal degree of freedom'' 
\cite{internalDOF}.
The recent work \cite{mulhern11} may be considered
as an extension of the above mentioned system class 
(i) to the case of two coupled underdamped particles.

In the context of nano-friction experiments 
by atomic force microscope (AFM), it has been reported \cite{soc06}
that an external time-periodic forcing may result in unmeasurably 
small sliding friction forces (so-called superlubricity).
Given that dimer models are well established to describe 
such systems \cite{goncalves04,maier05,goncalves05,tiwari08},
our present results predict the possibility of an even more 
astonishing effect, namely sliding friction with negative
sign, i.e. opposite to the direction of the pulling 
velocity.

For colloidal particles in magnetic or optical traps
\cite{magnetic,optic}, an effective ``bias'' force
is often generated by moving the traps with
respect to the ambient fluid and then going over 
into the comoving reference frame.
Along the lines of our present work, it might be 
possible to make colloidal doublets \cite{tie08}
move oppositely to the moving (and usually 
``entraining'') traps.

\begin{center}
\vspace{-5mm}
---------------------------
\vspace{-4mm}
\end{center}
This work was supported by Deutsche Forschungsgemeinschaft 
under RE1344/5-1 and SFB 613, and by the ESF program
FANAS (collaborative research project
Nanoparma-07-FANAS-FP-009, RE1344/6-1).

\end{document}